\documentclass[runningheads]{svmult}

\usepackage{makeidx}   % allows index generation
\usepackage{graphicx}  % standard LaTeX graphics tool
\usepackage{subeqnar}  % subnumbers individual equations
\usepackage{multicol}  % used for the two-column index
\usepackage{physprbb}  % modified textarea for proceedings,
\makeindex             % used for the subject index
\begin{document}
\title*{Clues on Galaxy and Cluster Formation from their Scaling Relations}
\toctitle{Clues on Galaxy and Cluster Formation from their Scaling Relations}
% allows explicit linebreak for the table of content
%
%
\titlerunning{Galaxy and Cluster Scaling Relations}
% allows abbreviation of title, if the full title is too long
% to fit in the running head
%
\author{Barbara Lanzoni\inst{1}
\and Luca Ciotti\inst{2}
\and Alberto Cappi\inst{1}
\and Giuseppe Tormen\inst{3}
\and Gianni Zamorani\inst{1}
}
\authorrunning{Barbara Lanzoni et al.}
% if there are more than two authors,
% please abbreviate author list for running head
%
%
\institute{INAF -- Osservatorio Astronomico di Bologna, via Ranzani 1, 40127,
Bologna, Italy  
\and Dip. di Astronomia, Universit\`a di Bologna, via Ranzani 1,
40127 Bologna, Italy  
\and Dip. di Astronomia, Universit\`a di Padova, vicolo
dell'Osservatorio 5, 35122 Padova, Italy 
}

\maketitle              % typesets the title of the contribution

\begin{abstract}
By means of high-resolution N-body simulations in a $\Lambda$CDM cosmology,
we verify that scaling relations similar to those observed for nearby galaxy
clusters are also defined by their dark matter hosts; the slopes, however,
are not the same. We then show that the scaling relations of galaxy clusters
can be explained as the result of the cosmological collapse of density
fluctuations at the appropriate scales, plus a systematic trend of the $M/L$
ratio with cluster mass. The empirical fact that the exponent of the
Faber-Jackson relation of elliptical galaxies is significantly different
(higher) than that of clusters, force us to conclude that the galaxy scaling
laws might derive from the cosmological collapse of density fluctuations at
the epoch when galactic scales became non-linear, plus modifications
afterward due to early-time dissipative merging.
\end{abstract}

\section{Observed and simulated scaling relations}
It has been shown \cite{S93,A98} that nearby galaxy clusters define scaling
relations involving their optical luminosity $L$, effective radius
$R_{\mathrm e}$ and velocity dispersion $\sigma$. Reanalyzing the sample of
\cite{S93}, we obtain $L\propto \sigma^{2.18\pm0.52}$ for the Faber-Jackson
(FJ) relation, $L\propto R_{\mathrm e}^{1.55\pm0.19}$ for the Kormendy
relation, and $L\propto R_{\mathrm e}^{0.9\pm 0.15} \,\sigma^{1.31\pm 0.22}$
for a nearly edge-on view of the Fundamental Plane, where $L$ is in
$10^{12}L_\odot/h^2$, $\sigma$ in 1000 km/s, $R_{\mathrm e}$ in Mpc$/h$.

Similar scaling laws are also followed by early-type galaxies, with the only
significant difference of the FJ relation: in fact, $L\propto \sigma^4$ for
luminous ellipticals \cite{Bern03a}.

For a sample of 13 massive ($M>10^{14}{\mathrm M}_\odot/h$) DM halos obtained
from N-body cosmological simulations \cite{num_proceed}, we have computed the
projected radius containing half the total mass ($ R_{\mathrm h}$) and the
mean projected velocity dispersion within it ($\sigma_{\mathrm h}$). Well
defined scaling relations relating $ R_{\mathrm h}$, $\sigma_{\mathrm h}$ and
the total mass $M$ are found, and are interpreted as the result of the
cosmological collapse of density fluctuations and the weak homology of DM
halos \cite{NFW}. The slopes of these relations, however, do not coincide
with the observed ones: the simplest way to satisfy at once the theoretical
and observational scaling laws is to assume that $M/L\propto L^{0.3}$, in
good agreement with what inferred observationally \cite{Girardi02}.

\begin{figure}
\begin{center}
\includegraphics[width=\textwidth]{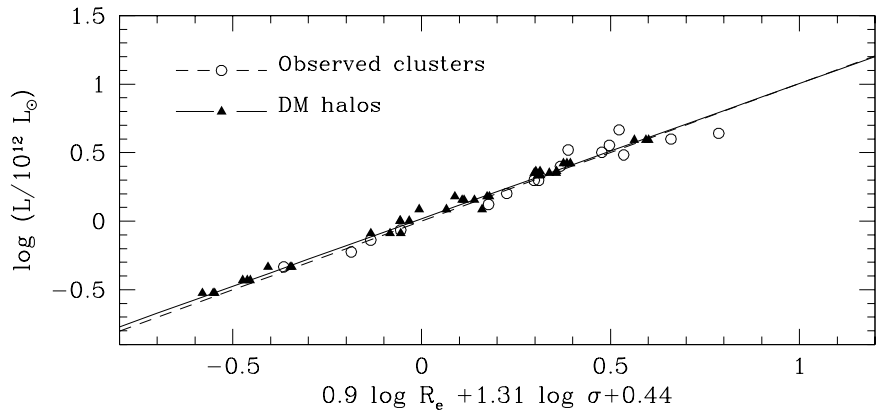}
\end{center}
\caption[]{Nearly edge-on view of the FP of galaxy clusters, and of their DM
halos under the assumption $M/L\propto L^{0.3}$}
\label{fig1}
\end{figure}

Since a similar trend of $M/L$ with $L$ is also found for early-type
galaxies, and the cosmological predictions for the DM halo scaling relations
are the same at galactic and cluster scales, the higher slope of the galactic
FJ relation implies that some additional mechanism played a role in the
formation/evolution of ellipticals: indeed, the combined effect of
dissipationless merging (which tends to increase the FJ exponent) and gas
dissipation (which acts in the opposite direction) might be the solution
\cite{fpcmer}.

\section{Conclusions}
We conclude that, while the scaling relations of galaxy clusters can be
explained by the cosmological collapse of density fluctuations at the
appropriate scale, plus a systematic trend the total $M/L$ ratio with cluster
luminosity, the scaling laws of elliptical galaxies seem to suggest that
additional processes (like early-time dissipational merging) probably played
an important role after the cosmological collapse of the density fluctuations
at these scales. For a discussion of how $M/L$ for clusters could increase
with $L$, we refer to \cite{fpcmer}.

%INDEX%%%%%%%%%%%%%%%%%%%%%%%%%%%%%%%%%%%%%%%%%%%%%%%%%%%%%%%%%%%%%%%
% Please check with the editor of your book whether he plans to
% include a "mutual" subject index - if so, please code your entries
% in the standard syntax. For your own purposes you may print your
% "personal" index by using the following commands:
%
%\clearpage
%\addcontentsline{toc}{section}{Index}
%\flushbottom
%\printindex
%%%%%%%%%%%%%%%%%%%%%%%%%%%%%%%%%%%%%%%%%%%%%%%%%%%%%%%%%%%%%%%%%%%%%

\end{document}